\documentclass[reprint,aps,prl,onecolumn,footinbib]{revtex4-1}
\usepackage{epsfig} 
\usepackage{amsmath}
\usepackage{amssymb}
\usepackage{mathrsfs}
\usepackage{amsthm}
\usepackage{enumerate}
\usepackage{siunitx}
\usepackage{graphicx}
\usepackage{bm}
\usepackage{fancyhdr}
\usepackage{textcomp}
\usepackage{gensymb}
\usepackage{hyperref}
\usepackage{float}
\usepackage{caption}
\usepackage{braket}
\usepackage{tikz}
\usepackage{xcolor}
\usepackage{tabularx}
\usetikzlibrary{quantikz}
\usepackage{booktabs}

\newcommand{\ra}[1]{\renewcommand{\arraystretch}{#1}}

\usepackage{lastpage}
\begin{document}

\setlength{\abovedisplayskip}{10pt}
\setlength{\belowdisplayskip}{10pt}

\title{Studies of the Fermi-Hubbard Model Using Quantum Computing}

\author{Adam Prokofiew, Nidhish Sharma, Steven Schnetzer}
\date{August 28, 2024}
\affiliation{\\ Rutgers University, Department of Physics and Astronomy}

\begin{abstract}
The use of quantum computers to calculate the ground state (lowest) energies of a spin lattice of electrons described by the Fermi-Hubbard model of great importance in condensed matter physics has been studied. The ability of quantum bits (qubits) to be in a superposition state allows quantum computers to perform certain calculations that are not possible with even the most powerful classical (digital) computers. This work has established a method for calculating the ground state energies of small lattices which should be scalable to larger lattices that cannot be calculated by classical computers. Half-filled lattices of sizes 1x4, 2x2, 2x4, and 3x4 were studied. The calculated energies for the 1x4 and 2x2 lattices without Coulomb repulsion between the electrons and for the 1x4 lattice with Coulomb repulsion agrees with the true energies to within 0.60\%, while for the 2x2 lattice with Coulomb repulsion the agreement is within 1.50\% For the 2x4 lattice, the true energy without Coulomb repulsion was found to agree within 0.18\%.

\end{abstract}

\maketitle 
\section{Introduction and Theory}
\raggedbottom
One of the exponentially difficult problems in physics and chemistry is the calculation of the states and energy levels of quantum systems. Simulation of systems with approximately 30 electrons or more is not feasible for classical computers. Quantum computers, however, are potentially much more powerful in solving these types of problems. Computational tasks on classical hardware that may require thousands of years of CPU time, especially ones that cannot be run in parallel, can be reduced to a reasonable processing time on the order of hours, minutes, or even seconds. This is due to the advantages of quantum computing including the ability to prepare qubits in a superposition of an exponential number of states.

The Hubbard Hamiltonian, a model known for its usefulness in condensed matter physics, is being investigated here within the quantum computing regime to first utilize an iterative optimization program to compute the minimum eigenvalue or equivalently the ground state energy of several simple electron lattice configurations. The longer term goal of this endeavor is to eventually use what is learned to scale up to more complex configurations that cannot be classically simulated. The Fermi-Hubbard Hamiltonian models electrons hopping across horizontally or vertically neighboring lattice sites.\cite{paper6} Diagonal hopping is not allowed. The model also accounts for Coulomb repulsion interactions between opposite spin electrons within each lattice site. Longer range interactions between different lattice sites are neglected. \\
\newline
\textbf{The Hubbard Hamiltonian}
\begin{equation}
H = -t \sum_{\langle i,j\rangle,\sigma} (a_{i\sigma}^{\dagger}a_{j\sigma}+a_{j\sigma}^{\dagger}a_{i\sigma})+U\sum_{k} n_{k\uparrow}n_{k\downarrow} -\mu\sum_{j}^{}(n_{j\uparrow}+n_{j\downarrow})
\end{equation} 
The first of the three terms in the Hubbard Hamiltonian, the kinetic energy, is called the hopping term. This term has a parameter $t$ which is the hopping amplitude; we set this equal to 1. The second term, the potential energy, is called the on-site term and has a parameter $U$, the strength of the Coulomb interaction. This investigation exclusively involves electrons, so $a_{i\sigma}^{\dagger}$ and $a_{i\sigma}$ are fermionic creation and annihilation operators, respectively, for neighboring lattice sites $i$ and $j$ of a given spin $\sigma \in \{\uparrow,\downarrow\}$. These operators must obey anti-commutation relations due to the Pauli exclusion principle. The close range Coulomb interaction of opposite spin at a given lattice site is represented by the terms $n_{k\uparrow} = a_{k\uparrow}^{\dagger}a_{k\uparrow}$ and $n_{k\downarrow} = a_{k\downarrow}^{\dagger}a_{k\downarrow}$. Although the creation and annihilation operators are not hermitian individually, the entire term within parenthesis of the first summation, $a_{i\sigma}^{\dagger}a_{j\sigma}+a_{j\sigma}^{\dagger}a_{i\sigma}$, is hermitian and is referred to as the hopping operator.\cite{quantumbook} 

The third term is the chemical potential, which is defined as the change in energy due to a change in the number of particles. In our model, we preserve the number of electrons of a given spin type, so we do not take this term into account. The \textbf{U} coefficient was taken to be either 0 or 1 in testing our quantum algorithms. The ratio of \textbf{U} to \textbf{T} is inversely proportional to temperature and can help us observe phase transitions.

A Jordan-Wigner mapping is used in which 2 qubits are required for each physical lattice site and each sequential physical qubit represents a lattice site for one spin type.\cite{paper2} For example, for a given spin in a lattice with 4 sites a state may be $\ket{1010}$ indicating that the first and third lattice sites are occupied by one electron each while the second and fourth sites are unoccupied. This same expression corresponds to the qubit state of 4 qubits, 2 of which are $\ket{1}$.
Since the Jordan-Wigner Transformation maps neighboring lattice sites to sequential qubits and hopping does not occur between electrons of opposite spin it is best to map the lattice sites of the second spin type sequentially following the final lattice site of the first spin type. I.e. opposite spins sites are sequentially segregated. See Fig. 1 below for a geometric example of this configuration.\cite{qiskitbook}

\begin{figure}[H] 
\centering
\includegraphics[width=.4\linewidth]{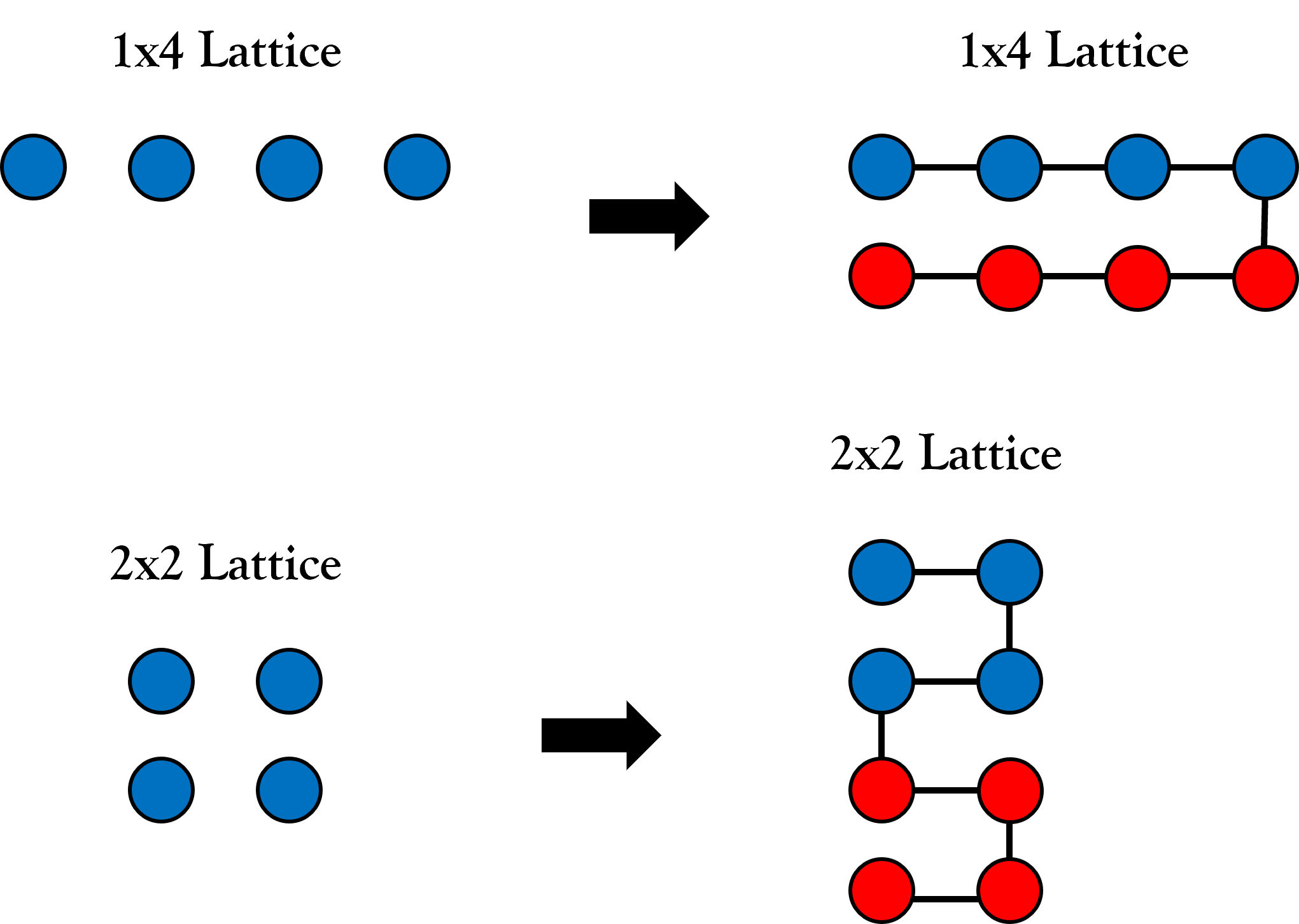}
\caption{Illustrated in this diagram are the geometric representations of the lattice configurations of this study. This diagram shows each lattice before and after mapping each physical lattice site to 2 qubits each(1 for each spin) via the Jordan-Wigner Transformation.}
\end{figure}
Starting from either mapped lattice on the right side of Fig. 1, the top left blue lattice site is the first qubit in the sequence and the black line represents the order of sequential qubits. Blue lattice sites represent one spin type and red the other spin type. The Coulomb interaction in the Hamiltonian is between electrons in corresponding lattice sites of opposite color. I.e. the first blue and the first red lattice sites along the sequence may interact, the second of each and so on.\\

\textbf{Variational Quantum Eigensolver}
\newline

Following the Jordan-Wigner transformation, a quantum system can be simulated using a variational quantum eigensolver. A variational quantum eigensolver is a hybrid quantum-classical computation algorithm that contains a parameterized quantum circuit. An initial superposition state is prepared. Within the ansatz, the section of the circuit that defines the parameterized state, initial parameters are chosen. The algorithm then optimizes the quantum circuit parameters iteratively with a classical optimizer over as many iterations as necessary in an attempt to extract the minimum eigenvalue of a given Hamiltonian. If possible, initial parameters should be chosen such that they are close to the global minimum. The expectation value of the Hamiltonian for the parameterized state at each iteration is the cost function to be minimized.\cite{rieffel}

\begin{figure}[H]
\centering
\includegraphics[width=0.9\linewidth]{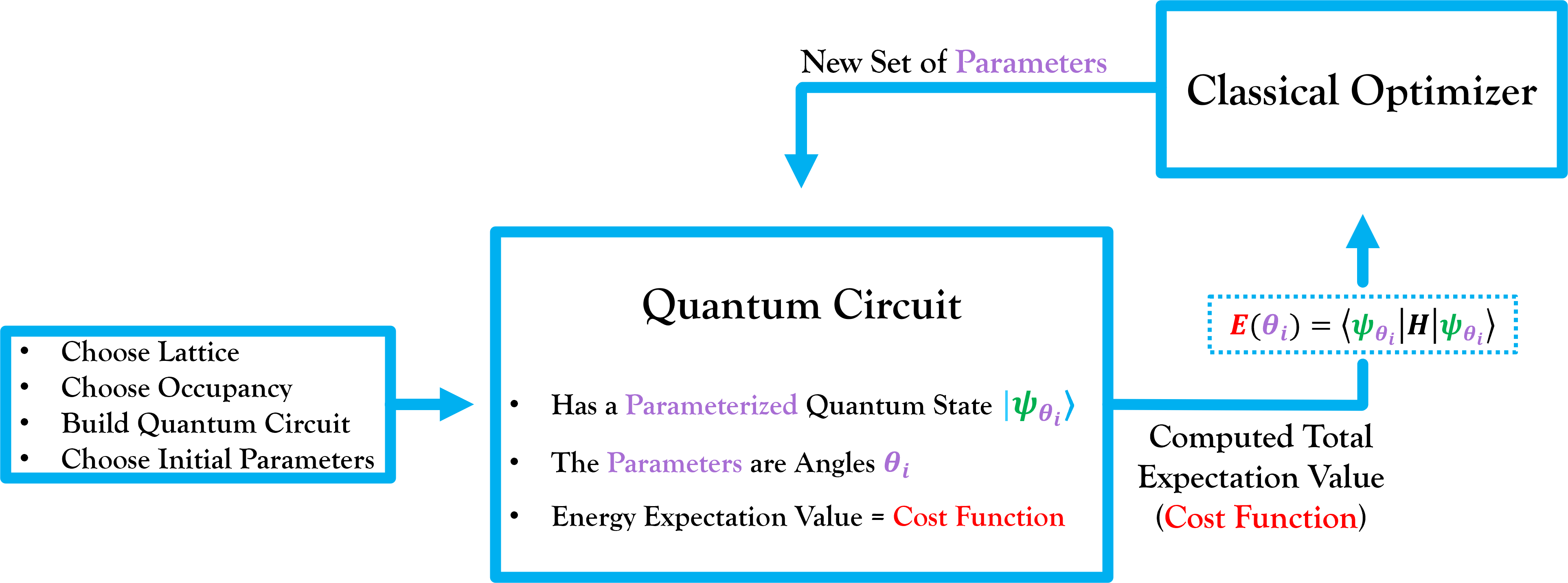}
\caption{This figure illustrates the general variational quantum eigensolver algorithm where the energy $E$ of a state $\ket{\psi_{\theta_i}}$ being optimized with parameters $\theta_i$ is equal to the expectation value of the Hamiltonian $\braket{H}$ as $E(\theta_i) = \bra{\psi_{\theta_i}}H\ket{\psi_{\theta_i}}$}
\end{figure}

\textbf{Quantum Circuit}\\

The model and mapping tools are combined to build a quantum circuit that consists of three main parts. 
The first part is an initialization circuit to set up the initial state such that it represents the physical system being simulated. This initial state is an ideally uniform superposition of as many of the physical states as possible, also referred to as physical basis states. Out of all possible binary combination states, physical states are a subcategory that preserve the electron occupancy for a given spin type. For example, consider both spin types of a 1x2 lattice that is half-occupied. It will have a single electron of each spin type or 2 total electrons and has the 4 physical basis states $\ket{0101}, \ket{0110}, \ket{1001}$, and $\ket{1010}$. Notice that $\ket{1100}$ and $\ket{0011}$ are not physical basis states despite preserving total electrons. This is because the first half of the bit string is one spin type, the second half is another, and spin type would not be preserved. As long as the number of electrons in each lattice site is equal and the lattice occupancy is half-filled, the corresponding state will be physical. 

The second part of the quantum circuit is called the ansatz, which is a series of quantum gate combinations that must preserve the number of electrons occupying each lattice of a given spin type and contain parameterized rotation gates. This is the section of the circuit that is essentially driven by an optimization function to search for the minimum energy using the parameters of the rotation gates. If necessary, the circuit depth of this ansatz can be increased to improve its ability to search by adding additional ansatz layers. A second layer will have an identical gate configuration, but the parameters of its rotation gates will be different to provide more degrees of freedom. 

The third and final part of the quantum circuit before measurement is the hopping circuit shown below in Fig. 3. The hopping circuit diagonalizes the hopping operator $a_{i\sigma}^{\dagger}a_{j\sigma}+a_{j\sigma}^{\dagger}a_{i\sigma}$ from the Hamiltonian in equation (1) before measurement of the expectation value for the kinetic energy term. A single hopping term along the path of the qubit mapping is calculated as $P_{i,j}(01)-P_{i,j}(10)$ where $P$ is the probability of lattice site occupancy for neighboring qubits $q_{i}$ and $q_{j}$ of the same spin type. The expectation value due to the on-site or Coulomb interaction is measured directly without a diagonalization circuit. A single on-site term is calculated as the probability $P_{i,k}(11)$ where qubit $q_{i}$ has on-site interaction correspondence to $q_{k}$ of opposite spin with $k=i+n$ where $n$ is the total number of physical lattice sites as shown below in equation (3). The total expectation value is the sum of the hopping terms and the onsite terms. It is this total expectation value that is the cost function to be minimized by the classical optimizer.\\

\textbf{Hopping Circuit}
\begin{figure}[H]
\centering
\includegraphics[width=0.5\linewidth]{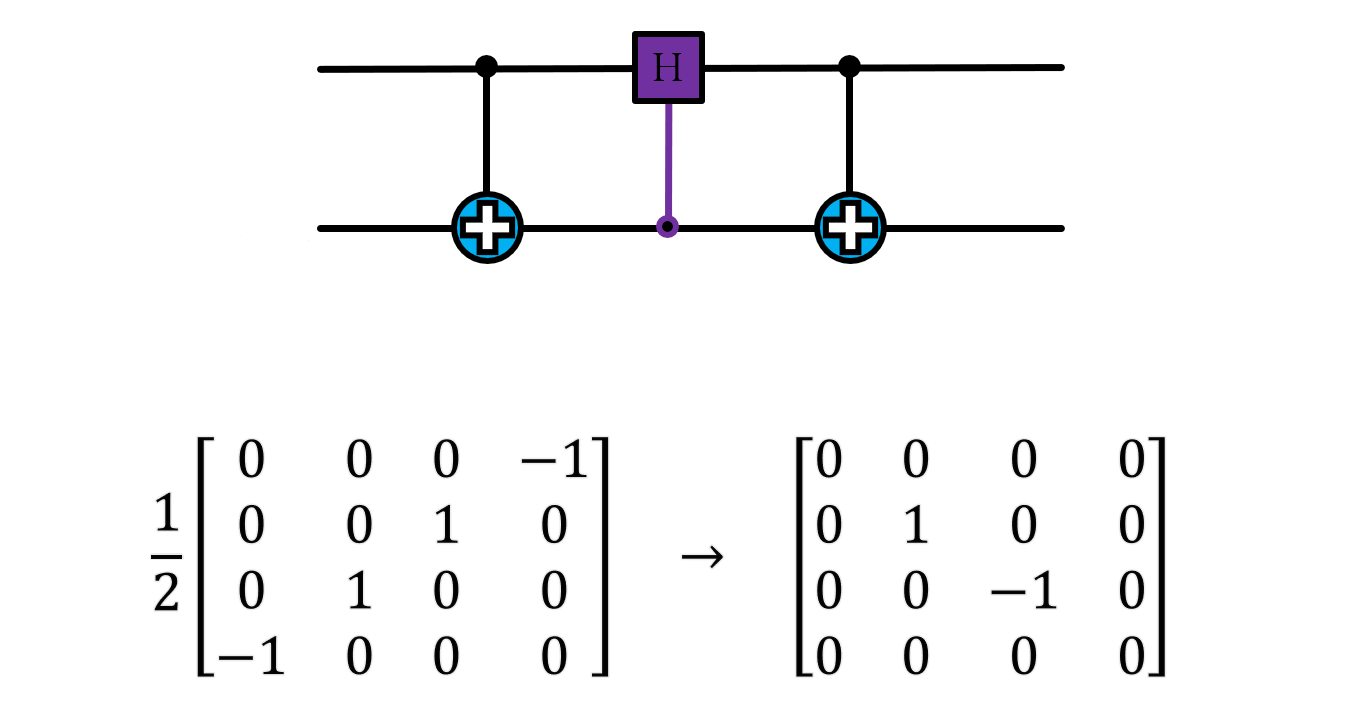}
\caption{This figure shows the hopping circuit CNOT$_{(q,q+1)}$ Control Hadamard$_{(q+1,q)}$ CNOT$_{(q,q+1)}$ where each gate has indices for some control qubit and target qubit, respectively. Below this circuit the figure shows hopping operator $a_{i\sigma}^{\dagger}a_{j\sigma}+a_{j\sigma}^{\dagger}a_{i\sigma}$ as a matrix before and after its diagonalization by the hopping circuit. The expectation value shown as $\braket{H}_{Hop}$ is the energy contribution or cost function contribution from the hopping terms.}
\end{figure}

\textbf{Hopping Term Expectation Value}
\begin{equation}
\braket{H}_{Hop} = -t \sum_{i,j}(P_{ij}(01)-P_{ij}(10))
\end{equation} 

\textbf{On-site Term Expectation Value}
\begin{equation}
\braket{H}_{Onsite} = U \sum_{i}P_{ik}(11)\;\;\;\;\;\;\;\;\;k=i+n
\end{equation}

\textbf{Total Expectation Value}
\begin{equation}
\braket{H} =  \braket{H}_{Hop}+\braket{H}_{Onsite}
\end{equation}

\section{Procedure}
The first step of the procedure is to choose a lattice shape and level of electron occupancy. Lattices "half-filled" or equivalently, lattices of $\frac{1}{2}$ occupancy were chosen for this study which are particularly interesting since many phenomena, such as Mott insulators, in condensed-matter physics are observable in this case.\cite{paper4} For small lattices, the theoretical ground state and its energy can be calculated to compare against the simulated results.\cite{paper1}

Once a lattice size and occupancy level is chosen, a simulated quantum circuit is built to the specifications and constraints of that chosen system as described in the $\mathbf{Quantum}$ $\mathbf{Circuit}$ sub-section above. This study will investigate the 1x4, 2x2, and 2x4 lattice configurations for the half-filled case. The first two configurations require 8 qubits total, 4 qubits for each spin type, and utilize identical initialization circuits for each spin type. The last configuration requires 16 qubits, with 8 for each lattice site. A block diagram for the quantum circuit for the 2x2 lattice is shown below in Fig. 4. Excluding the third hopping circuits labeled $\textcolor{red}{Hop*}$ from the block diagram gives the general form of the 1x4 lattice. The final component of the circuit is measurement as shown in Fig. 4 by the small meter symbols at the end of each circuit line where each qubit is measured.
The 1x4 or 2x2 initialization circuit shown below in Fig. 5 creates a superposition of its 6 half-filled physical basis states $\ket{0011}$, $\ket{0101}$, $\ket{0110}$, $\ket{1001}$, $\ket{1010}$, and $\ket{1100}$ when considering one spin type individually. When considering the entire quantum circuit including both spin types (which is necessary after they become entangled), there are always 36 physical basis states beginning as the tensor product (before entanglement) of both 6 physical basis state circuits of each spin type. \\
\newpage
\textbf{2x2 or 1x4* Lattice Quantum Circuit}
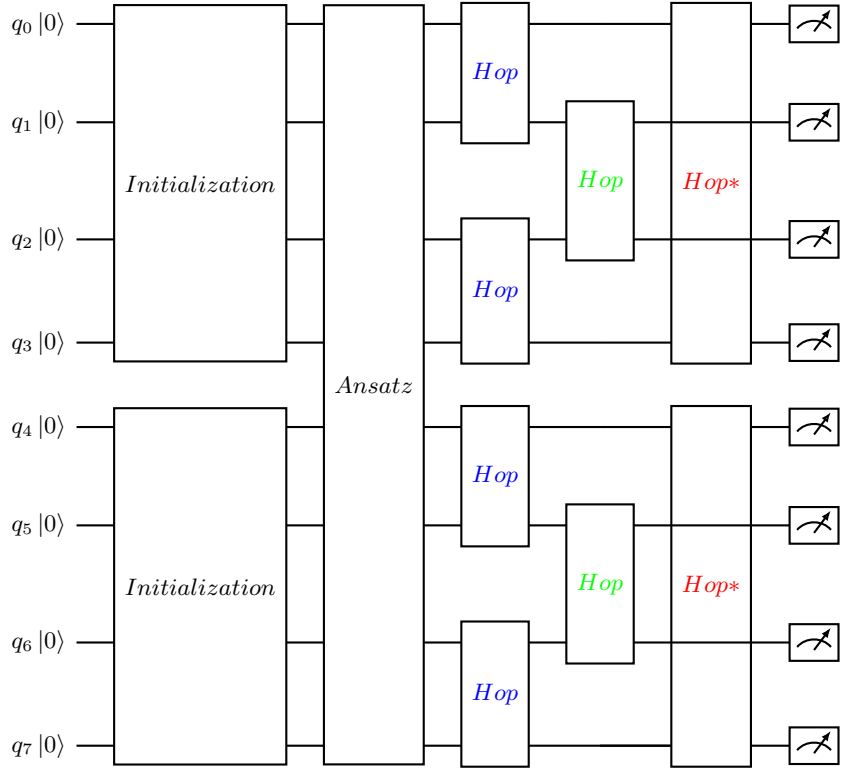
\begin{figure}[H]
\centering
\begin{quantikz}[transparent]
\lstick{$q_{0}\ket{0}$} & \gate[4]{Initialization} & \gate[8]{Ansatz} & \gate[2]{\color{blue}Hop} & \qw & \gate[4]{\color{red}Hop*} & \meter{} \\
\lstick{$q_{1}\ket{0}$} & & & & \gate[2]{\color{green}Hop} & \linethrough & \meter{} \\
\lstick{$q_{2}\ket{0}$} & & & \gate[2]{\color{blue}Hop} & & \linethrough & \meter{} \\
\lstick{$q_{3}\ket{0}$} & & & & \qw & & \meter{} \\
\lstick{$q_{4}\ket{0}$} & \gate[4]{Initialization} & & \gate[2]{\color{blue}Hop} & \qw & \gate[4]{\color{red}Hop*} & \meter{} \\
\lstick{$q_{5}\ket{0}$} & & & & \gate[2]{\color{green}Hop} & \linethrough & \meter{} \\
\lstick{$q_{6}\ket{0}$} & & & \gate[2]{\color{blue}Hop} & & \linethrough & \meter{} \\
\lstick{$q_{7}\ket{0}$} & & & & \qw & \qw & \meter{} \\
\end{quantikz}
\caption{This figure illustrates the general form of the quantum circuits used within the variational quantum eigensolver for a physical lattice size of 4 sites. Hopping circuits that commute (i.e. can be measured on a single circuit) are color coded. *Notice the hopping circuits labeled $\textcolor{red}{Hop*}$ between $q_{0}$ $\leftrightarrow$ $q_{3}$ and $q_{4}$ $\leftrightarrow$ $q_{7}$. These non-sequential qubit vertical lattice hopping circuits only apply to a 2x2 lattice and not a 1x4 lattice where the lattice sites are too far from each other for allowed hopping.}
\end{figure}

\textbf{1x4 or 2x2 Initialization}
\begin{figure}[H]
\centering
\includegraphics[width=0.32\linewidth]{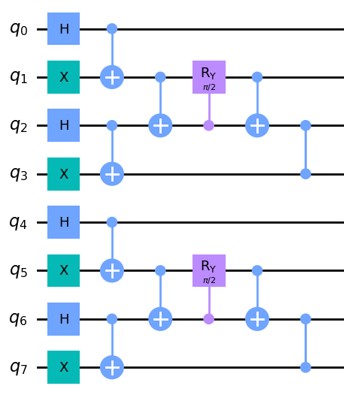}
\caption{Shown in this figure is the initialization for half-filled occupancy of the 1x4 or 2x2 lattice configuration giving each spin type the initial state $\ket{\psi_{0}} = \frac{\sqrt{2}}{4}\ket{0011}+\frac{\sqrt{2}}{4}\ket{0101}+\frac{1}{2}\ket{0110}+\frac{1}{2}\ket{1001}+\frac{\sqrt{2}}{4}\ket{1010}+\frac{\sqrt{2}}{4}\ket{1100}$ and giving the initial state for both spins as tensor product ${\ket{\psi_{0}}}\otimes\ket{\psi_{0}}=\ket{\psi_{0}\psi_{0}}$ of 36 basis states. This circuit contains two CRY gates which are controlled Y-axis rotation gates. In this initialization their angles are chosen as $\frac{\pi}{2}$ to prepare the desired initial state. }
\end{figure}

Shown below in Fig. 6 is the 2x2 lattice ansatz. The design of the ansatz is constrained. It must preserve the half-filled electron occupancy for each spin type. The ansatz is partitioned into three sections by gray barriers for organization. The first section containing 4 CRZ gates (controlled Z-axis rotation) is responsible for allowing communication between opposite spin types when Coulomb repulsion is nonzero (i.e., when $U > 0$ opposite spin qubits and circuits become entangled). Each pair of qubits that correspond to the same physical lattice site for Coulomb repulsion are linked by a single CRZ gate.\cite{paper5} The second section consists of commutative 4-gate combinations containing CNOT (controlled "NOT" or bit flip), CRX (controlled X-axis rotation), and CRZ gates that enable communication between sequential qubits of the same spin type for lattice hopping. One of each of these gate combinations is implemented in the second section for each pair of sequential qubits. The third section contains the final set of parameters responsible for communication between $q_{0}$ $\leftrightarrow$ $q_{3}$ and between $q_{4}$ $\leftrightarrow$ $q_{7}$. This third section is needed for the 2x2 lattice which has hopping between $q_{0}$ $\leftrightarrow$ $q_{3}$ and between $q_{4}$ $\leftrightarrow$ $q_{7}$. \\

\textbf{2x2 Ansatz}
\begin{figure}[H]
\centering
\includegraphics[width=0.8\linewidth]{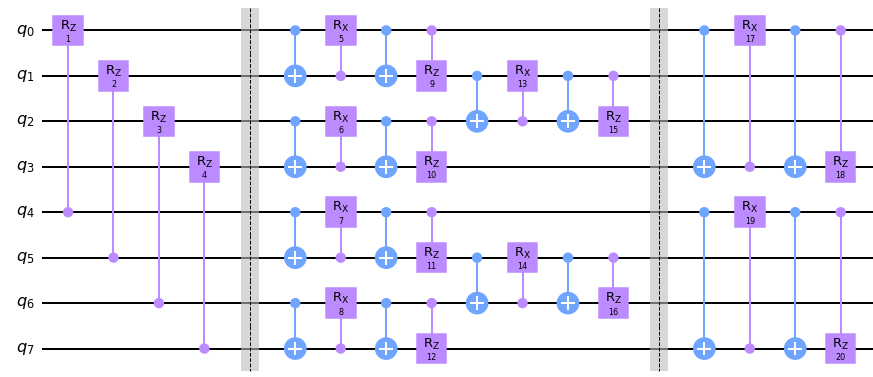}
\caption{Shown in this figure is the ansatz for the 2x2 lattice configuration. The 1x4 ansatz is represented here also by simply excluding the third section of this circuit following the second gray barrier containing control gates communicating between $q_{0}$ $\leftrightarrow$ $q_{3}$ and also between $q_{4}$ $\leftrightarrow$ $q_{7}$ for the second spin type. The 2x2 ansatz requires these extra gates due to non-sequential qubit vertical lattice hopping.}
\end{figure}

Shown below are the three hopping circuits that do not commute with each other and thus must be measured on separate circuits referred to as Hopping Circuit 1, 2, and 3 below in Fig. 7, Fig. 8, and Fig. 9, respectively. These figures correspond to $\textcolor{blue}{Hop}$, $\textcolor{green}{Hop}$, and $\textcolor{red}{Hop*}$, respectively, as shown in the quantum circuit block diagram in Fig. 4.\\

\textbf{1x4 or 2x2 Hopping Circuit 1}
\begin{figure}[H]
\centering
\includegraphics[width=.5\linewidth]{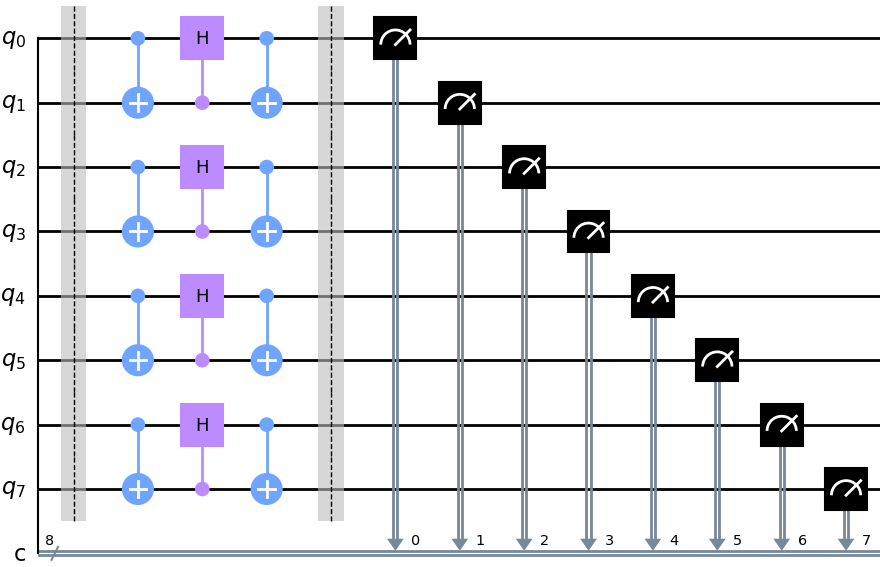}
\caption{Shown in this figure is the first set of hopping circuits for both spins types that correspond to the blue \textcolor{blue}{$Hop$} circuits in Fig. 4.}
\end{figure}

\textbf{1x4 or 2x2 Hopping Circuit 2}
\begin{figure}[H]
\centering
\includegraphics[width=.5\linewidth]{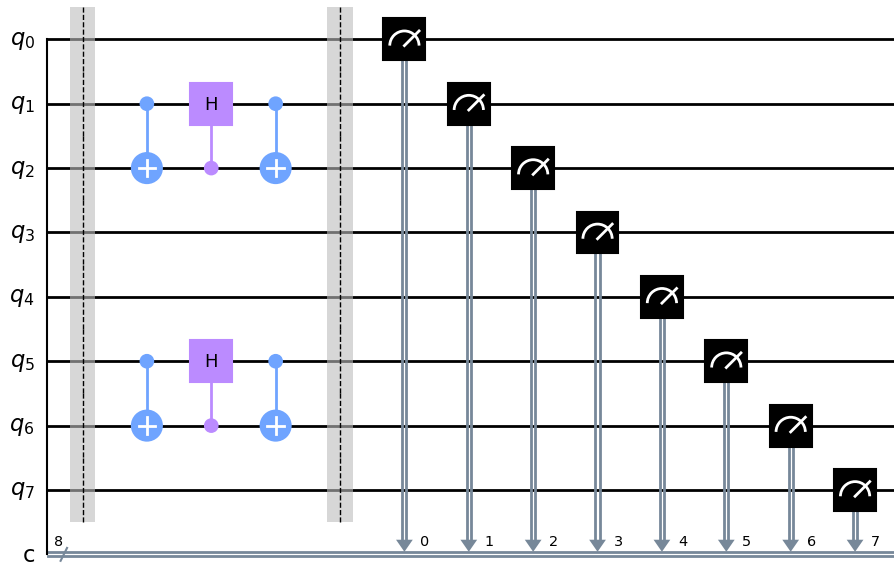}
\caption{Shown in this figure is the second set of hopping circuits for both spin types that correspond to the green \textcolor{green}{$Hop$} circuits in Fig. 4.}
\end{figure}

The Pauli exclusion principle must be obeyed by electrons, therefore special consideration must be given for use of the hopping circuit between non-sequential qubits (i.e. hopping not along the Jordan-Wigner mapping). Since the half-filled case is the only occupancy included in this study, the "standard" hopping circuit 3 shown below in Fig. 9 for sequential qubits can be used for the non-sequential qubit hopping of the 2x2 lattice (between $q_{0}$ $\leftrightarrow$ $q_{3}$ and between $q_{4}$ $\leftrightarrow$ $q_{7}$) with the simple modification of applying a minus sign to that particular term's probability contribution inside the cost function to maintain parity. More generally, use of fermionic swap gates (FSWAP) are required for other occupancy levels and are configured within the circuit shown below in Fig. 10. A single FSWAP gate consists of a control Z gate and SWAP gate as a combination. This can be seen below in Fig. 10 where 8 FSWAPs are shown total, 4 surrounding each "standard" or sequential hopping circuit of each spin type. \\

\textbf{2x2 Hopping Circuit 3 For Half-Filled Case*}
\begin{figure}[H]
\centering
\includegraphics[width=.5\linewidth]{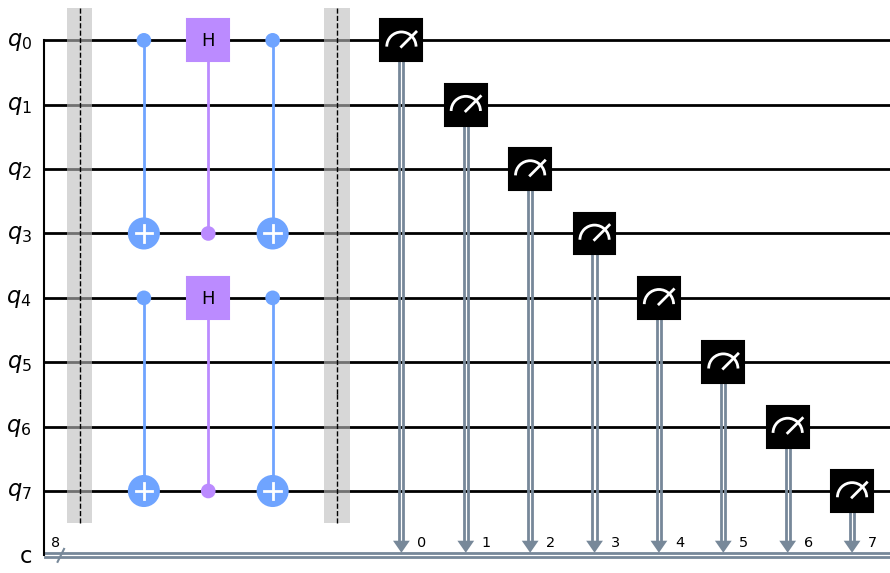}
\caption{Shown in this figure is the third set of hopping circuits for both spin types that correspond to the red \textcolor{red}{$Hop*$} circuits in Fig. 4 that represents non-sequential qubit vertical lattice hopping for nonlinear lattice configurations such as the 2x2. Additionally, when this hop between $q_{0}$ $\leftrightarrow$ $q_{3}$ and $q_{4}$ $\leftrightarrow$ $q_{7}$ is not for the $\frac{1}{2}$ occupancy or half-filled case, a different hopping circuit configuration is required using gates called fermionic swap gates or "FSWAP" gates. }
\end{figure}

\newpage

\textbf{2x2 Hopping Circuit 3 For Any Occupancy Case}
\begin{figure}[H]
\centering
\includegraphics[width=.5\linewidth]{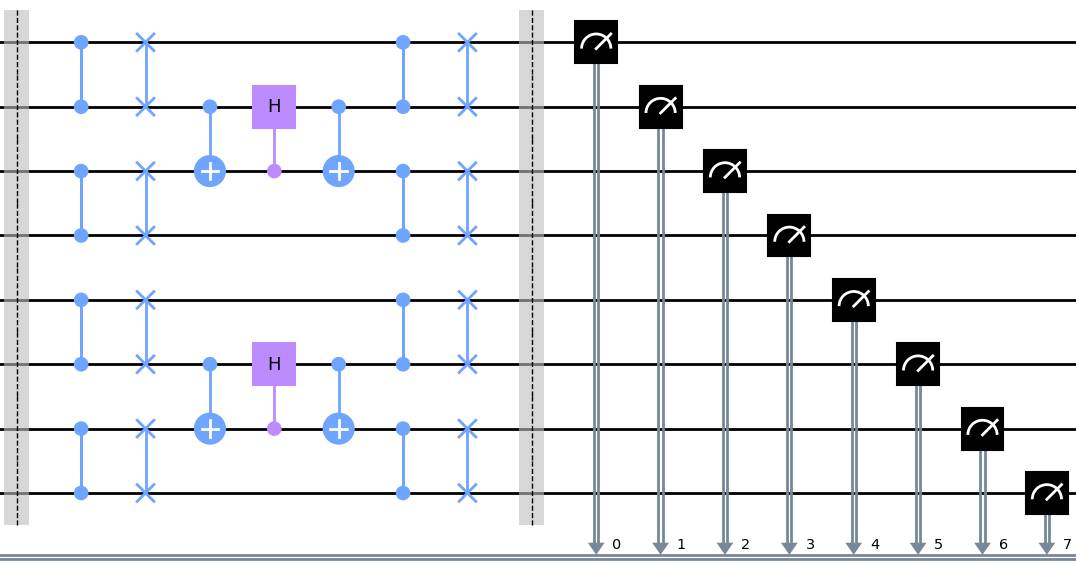}
\caption{Shown in this figure is the generalized 2x2 hopping circuit for non-sequential qubit hopping using FSWAP gates to diagonalize hopping between $q_{0}$ $\leftrightarrow$ $q_{3}$ as well as $q_{4}$ $\leftrightarrow$ $q_{7}$ before measurement.}
\end{figure}

The FSWAP accounts for the Pauli exclusion anti-symmetry of electrons and enables the circuit to maintain parity by swapping the necessary number of adjacent qubits with one sequential swap at a time such that the non-sequential qubits are swapped into the position of the Jordan-Wigner qubit sequence.\cite{paper3} The hopping circuit can then operate on initially non-sequential qubits (such as $q_{0}$ and $q_{3}$ in the 2x2) using the sequential qubit hopping circuit and then swap them back before measurement as shown above in Fig. 10. For larger nonlinear lattices such as a 2x3, additional FSWAPs before and after the hopping circuit is required for its outer-most non-sequential qubits to be diagonalized before measurement. The corresponding FSWAP matrix is given by 
\begin{equation*}
\begin{bmatrix}
1 & 0 & 0 & 0 \\
0 & 0 & 1 & 0 \\
0 & 1 & 0 & 0 \\
0 & 0 & 0 & -1
\end{bmatrix}
\end{equation*}

\textbf{2x4 Lattice}\\

The largest lattice size considered is the 2x4 case. Similar to the previous 2x2 and 1x4 cases, we can simply extend parts of the algorithm to account for the additional qubits. Figure 11 shows the resulting ansatz that allows for communication between qubits representing neighboring lattice sites. Higher lattice sites will follow the same qubit communication pattern. 

\begin{figure}[ht!]
    \centering
    \includegraphics[width=0.29\linewidth]{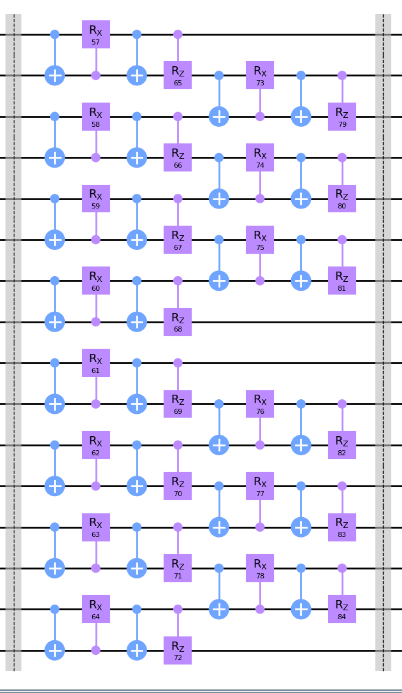}
    \caption{2x4 Small Ansatz extended from the previous lattice sizes. This is repeated a number of times depending on how many layers the corresponding run uses}
    \label{fig:enter-label}
\end{figure}
\newpage

The initialization is similarly extended for the 2x4 case. The possible basis states for an arbitrary number of qubits can be expressed with the following:
\begin{equation}
    \ket{\psi} = \sum_{i=0}^{2^{n}-1}\frac{1}{2^{n/2}}x_{i}\ket{bin(i)}
\end{equation}

\textit{bin(i)} corresponds to the binary representation of the corresponding index in the sum. This matches with Qiskit's notation for qubit expression. The circuit output can be seen below in figure 12.

\begin{figure}[ht!]
    \centering
    \includegraphics[width=0.5\linewidth]{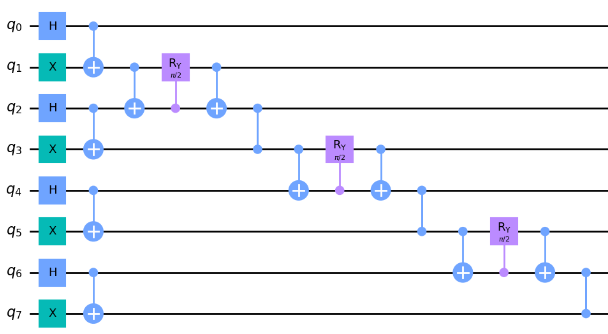}
    \caption{Initialization Circuit for the 2x4 Case. The same procedure is extended from the 2x2 case}
    \label{fig:enter-label}
\end{figure}

Figure 12 displays the circuit component which outputs physical states. For higher lattice sites, this procedure may not encompass all possible physical basis states so the implementation of Qiskit's statevector \textit{initialize} feature was discussed. This would allow us to assign an equal probability amplitude to each specific basis state and therefore make a completely full set of basis states to describe the lattice. However, this feature was incompatible with our GPU trials. This is a current avenue of investigation for improving the accuracy of our optimizer.\newpage

\begin{figure}[ht!]
    \centering
    \includegraphics[width=0.45\linewidth]{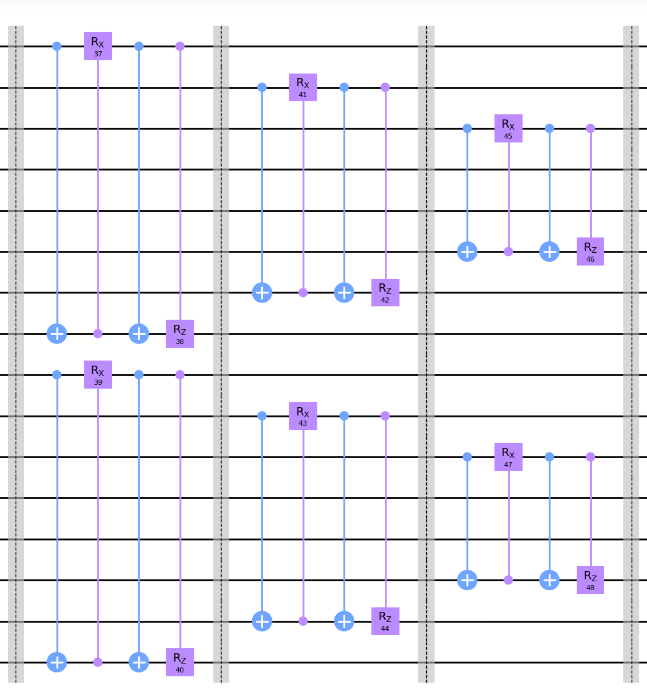}
    \caption{Hopping circuits for the 2x4 case. An increase in qubits increases the number of hopping circuits needed}
    \label{fig:enter-label}
\end{figure}

Figure 13 displays the gates needed to enable communication between lattice sites and preserves the number of electrons occupying each site. In this figure, communication is enabled between qubits 0-7, 1-6, and 2-5. For reference, we can look at Figure 14 below:
\begin{figure}[ht!]
    \centering
    \includegraphics[width=0.5\linewidth]{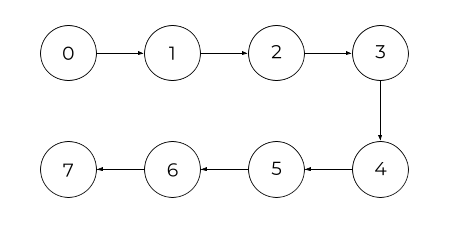}
    \caption{Visual Representation of 2x4 Lattice - Site Communication}
    \label{fig:enter-label}
\end{figure}

The final part of the circuit accounts for the hopping of non-sequential qubits. The list of these circuits is given below where the numbers refer to the qubit indices. 

\begin{enumerate}
    \centering
    \item Hopping Circuit 2: 1-2,3-4,5-6 
    \item Hopping Circuit 3: 0-7
    \item Hopping Circuit 4: 1-6
    \item Hopping Circuit 5: 2-5
\end{enumerate}

This list corresponds to the spin-up lattice; the spin-down lattice would be the same with each qubit index increasing by 8. Hopping Circuits 3 and 4 require an FSWAP which swaps the qubits to a lattice site where they are sequential, performs the operation, and then swaps them back while maintaining parity.

\section{data and analysis}

The classical optimizer used for this study was SPSA (Simultaneous Perturbation Stochastic Approximation). Due to the stochastic nature of the process, the ground state energy $E_{Simulated}$ optimization results vary run to run (a single "run" is shown by figures 15-19 where each figure is a single optimization run). In order to obtain accurate results within this varying distribution of results, 20 individual optimization runs, each with many iterations, were taken for each configuration and the best (lowest) result out of 20 was chosen (this can be done since the lowest possible eigenvalue is the ground state energy and therefore the lowest result must be the most accurate). Each result for a single run is taken as an average of the energy expectation values $\braket{H}$ of the last 100 iterations of optimization to account for the continuously fluctuating nature of the SPSA optimizer. Additionally, every probability (from expectation values) calculated in the process of obtaining results was computed using 40,000 shots to ensure accurate probability contributions for measurement. Hopping term parameter $t$ is constant as $t=1$, Coulomb repulsion or on-site parameter $U$ was varied as either $U=0$ or $U=1$, and energy values are considered with arbitrary energy units. The best simulated ground state energy $E_{Simulated}$ results are shown below in Table I. The true ground state energies shown below in Table I that are used for comparison against simulated ground state energy results were calculated by matrix calculations using a classical computer.  \\
\newpage
\textbf{Simulated Ground State Energy Best Results}
\begin{table}[H]
\centering
\ra{ 1.3 }
\begin{tabular}{|c|c|c|c|c|c|c|c|} 
\toprule
Lattice & $U$ & True $E_{g}$ & Best $E_{Simulated}$ & \%     Error & Ansatz Layers & Iterations & Size\\
\midrule
1x4 & 0 & -2$\sqrt{5}$ & -4.4707 & 0.03\% & 2 & 2000 & Small\\
1x4 & 1 & -3.5753 & -3.5547 & 0.58\% & 3 & 2000 & Small\\
2x2 & 0 & -4.00000 & -3.9969 & 0.08\% & 1 & 1000 & Small\\
2x2 & 1 & -3.3408 & -3.2918 & 1.47\% & 4 & 6000 & Small\\
2x4 & 0 & -10.47 & -10.4500 & 0.18\% & 3 & 50000 & Large\\
2x4 & 1 & -8.63 & -7.834 & 9.22\% & 4 & 86000 & Small\\
\bottomrule
\end{tabular}
\caption{Ground state energy $E_{Simulated}$ best results of studied lattices for different values of $U$. Hopping amplitude $t$ = 1 for all configurations. Results shown are best out of 20 SPSA optimization runs for each case.}
\end{table}

The first result in Table I is shown below in Fig. 15 below where the 1x4 lattice with U=0 (No Coulomb repulsion) has a true ground state energy of $E_{g}=-2\sqrt{5}$. The simulated result $E_{Simulated}=-4.4707$ is in agreement with the true energy at approximately 0.03\% error. Two ansatz layers were implemented to produce this result, as a single ansatz layer was not sufficient to reach close agreement of the true ground state energy. \\

\textbf{1x4 Lattice U=0 Best Result}
\begin{figure}[H]
\centering
\includegraphics[width=.7\linewidth]{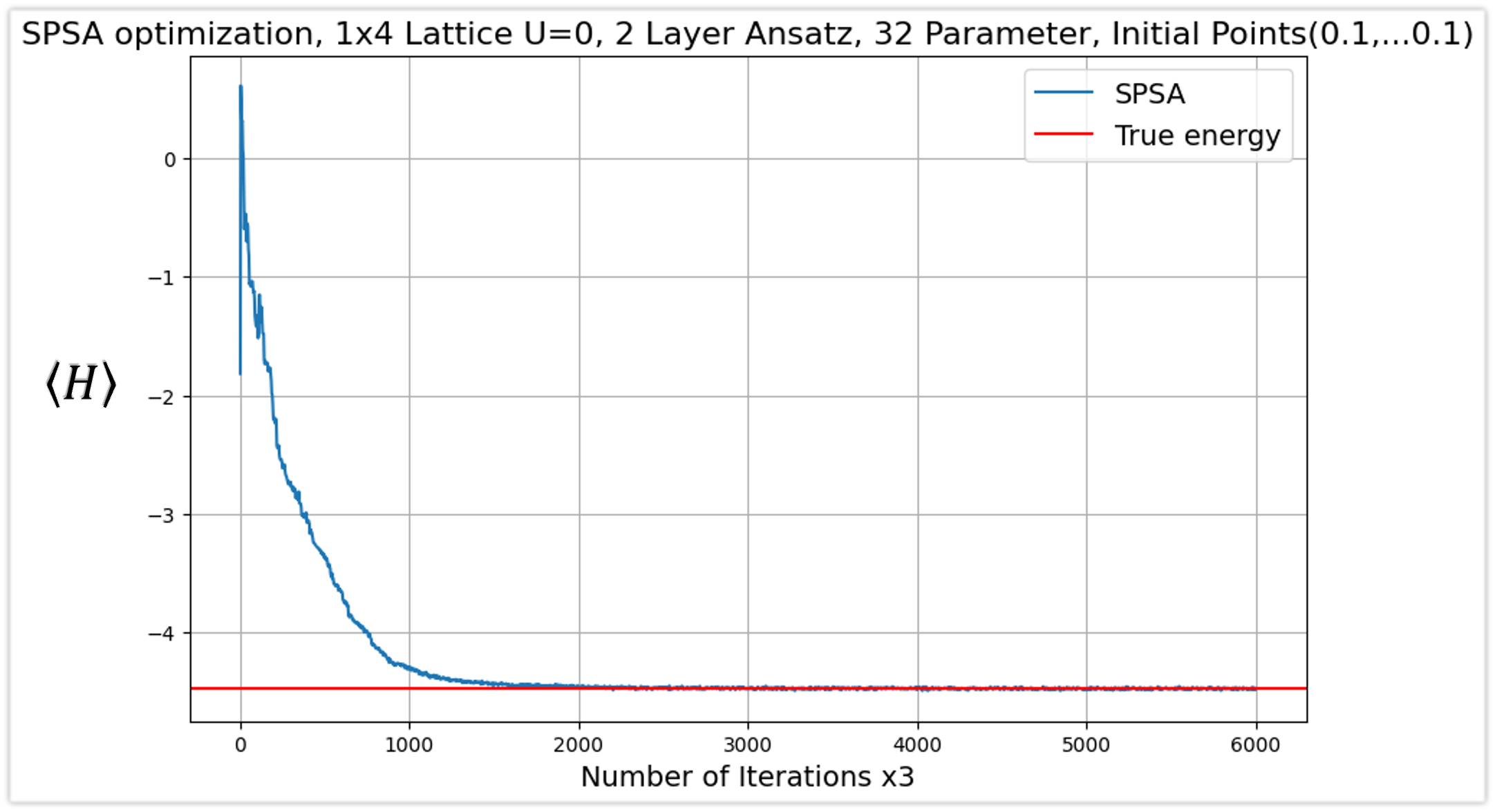}
\caption{This figure illustrates SPSA iteratively computing and converging to the true ground state energy for the 1x4 U=0 case. This was done using 2 Ansatz Layers of the maximum number of 16 parameters possible per layer for the 1x4 Ansatz.}
\end{figure}

The second result in Table I is shown below in Fig. 16 below where the 1x4 lattice with U=1 (Coulomb repulsion included) has a true ground state energy of approximately $E_{g}=-3.5753$. The simulated result of $E_{Simulated}=-3.5547$ is in agreement with the true energy at approximately 0.58\% error. Three ansatz layers were implemented in order to produce this result, as one or two ansatz layers was not sufficient to reach within close agreement of the true ground state energy. \\

\textbf{1x4 Lattice U=1 Best Result}
\begin{figure}[H]
\centering
\includegraphics[width=.7\linewidth]{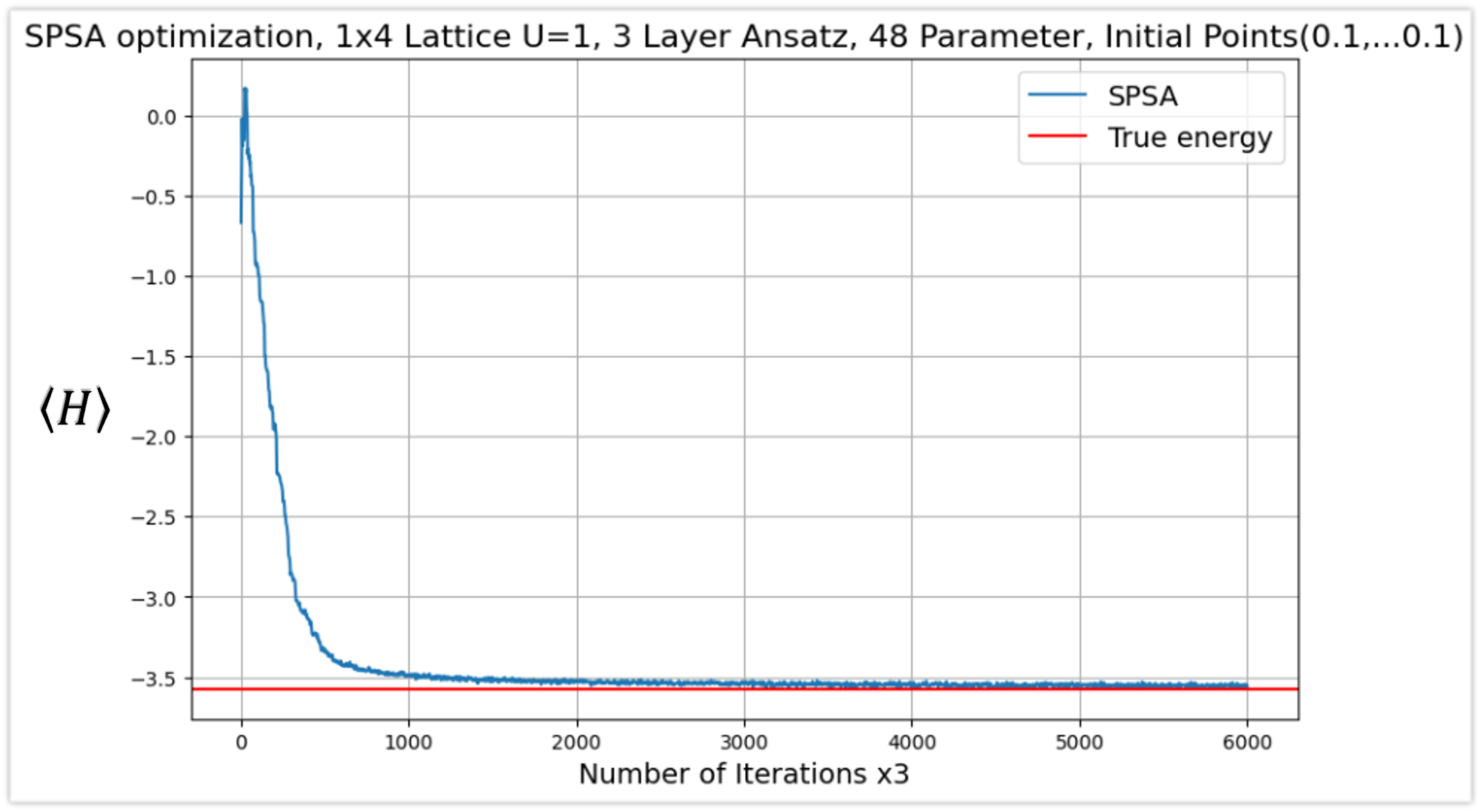}
\caption{This figure illustrates SPSA iteratively computing and converging to the true ground state energy for the 1x4 U=1 case. This was done using 3 Ansatz Layers of the maximum number of 16 parameters possible per layer for the 1x4 Ansatz.}
\end{figure}

The third result in Table I is shown below in Fig. 17 below where the 2x2 lattice with U=0 (No Coulomb repulsion) has a true ground state energy of approximately $E_{g}=-4.00000$. The simulated result of $E_{Simulated}=-3.99697$ is in agreement with the true energy at approximately 0.08\% error. A single ansatz layer was sufficient for this result. \\

\textbf{2x2 Lattice U=0 Best Result}
\begin{figure}[H]
\centering
\includegraphics[width=.7\linewidth]{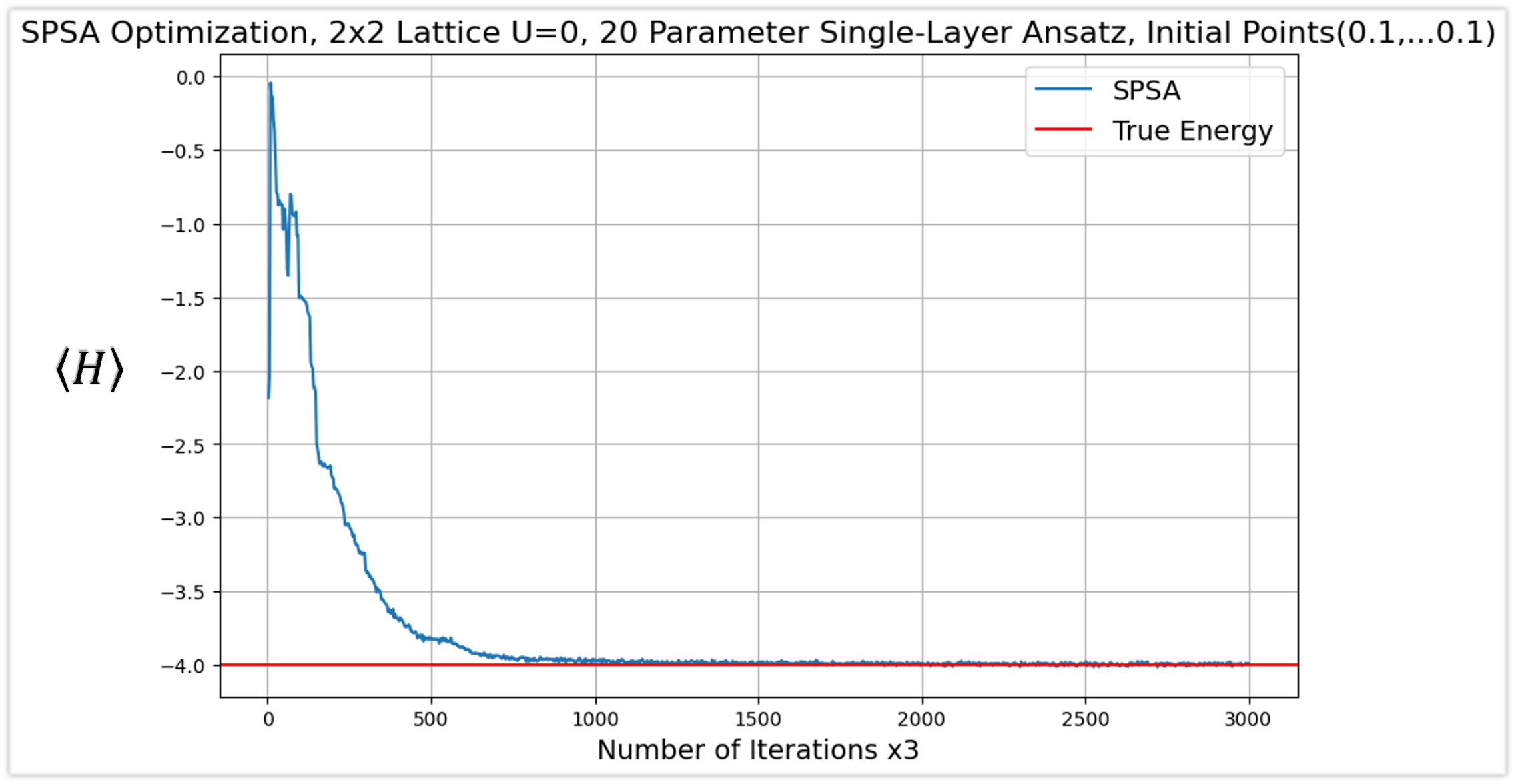}
\caption{This figure illustrates SPSA iteratively computing and converging to the true ground state energy for the 2x2 U=0 case. This was done using 1 Ansatz Layer of the maximum number of 20 parameters possible per layer for the 2x2 Ansatz.}
\end{figure}

The fourth result in Table I is shown below in Fig. 18 below where the 2x2 lattice with U=1 (Coulomb repulsion included) has a true ground state energy of approximately $E_{g}=-3.3408$. The simulated result of $E_{Simulated}=-3.2918$ is in agreement with the true energy at approximately 1.47\% error. Four ansatz layers were implemented in order to produce this result, as one, two, or three ansatz layer configurations were not sufficient to reach within similar agreement of the true ground state energy as results for other cases. \\

\newpage
\textbf{2x2 Lattice U=1 Best Result}
\begin{figure}[H]
\centering
\includegraphics[width=.7\linewidth]{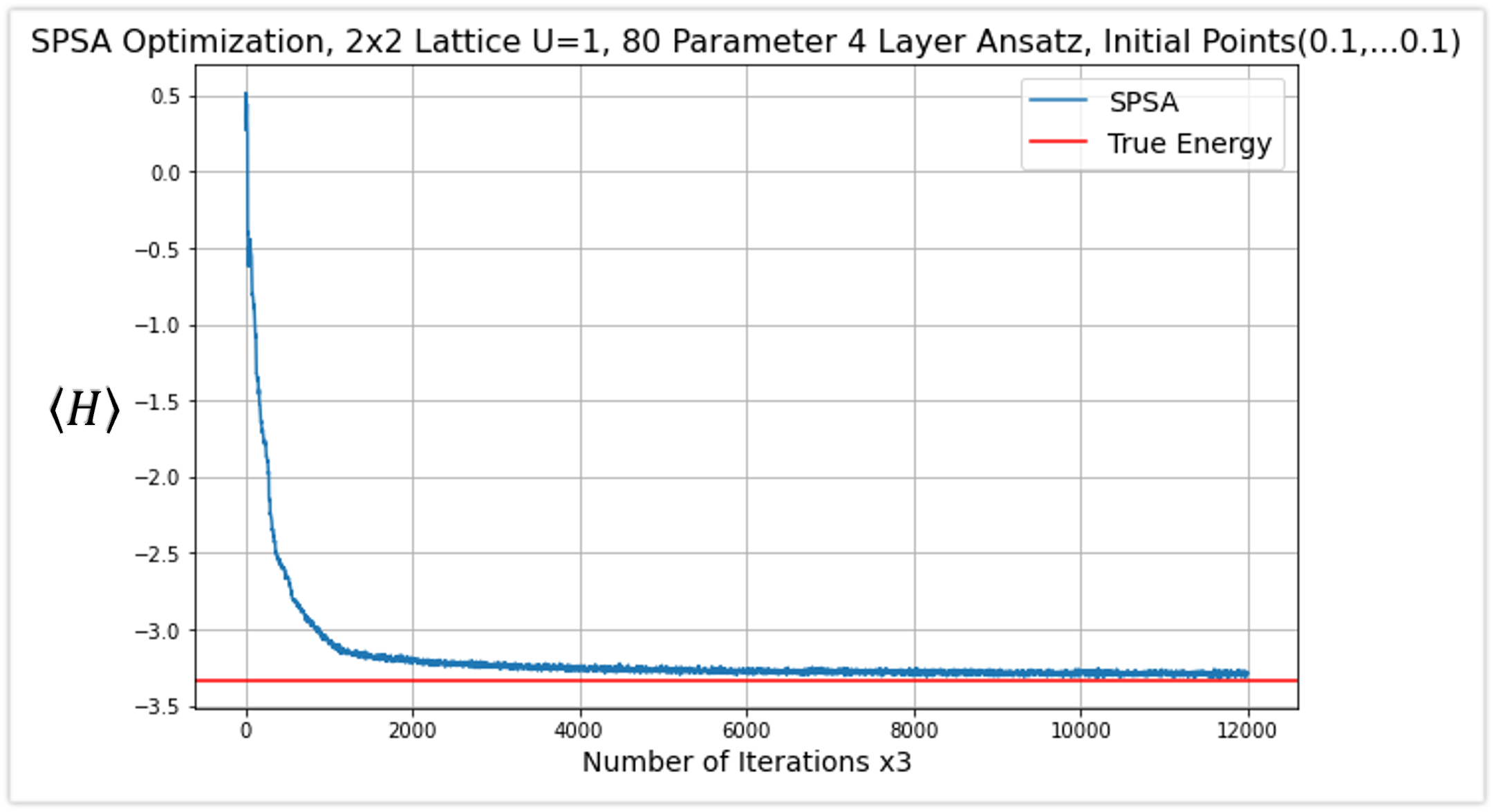}
\caption{This figure illustrates SPSA iteratively computing and converging to the true ground state energy for the 2x2 U=1 case. This was done using 4 Ansatz Layers of the maximum number of 20 parameters possible per layer for the 2x2 Ansatz.}
\end{figure}

\textbf{2x4 Lattice U=0 Best Results}\\

Figure 19 displays the SPSA optimization for the 5-layer big ansatz trial case. The energy difference was 0.028 with a percent error of 0.25\%. This mirrors the behavior of the 3-layer result recorded in table 1. The ground state energy is $E_g=-10.47$ and the simulated result was 10.442. Up to five ansatz layers were implemented in order to see the overall effect of the number of layers on the accuracy of the result. Evidently, increasing the number of layers from 3 to 5 does not give a significantly better result which indicates a cutoff in the increase of the accuracy. \\

\begin{figure}[ht!]
    \centering
    \includegraphics[width=0.7\linewidth]{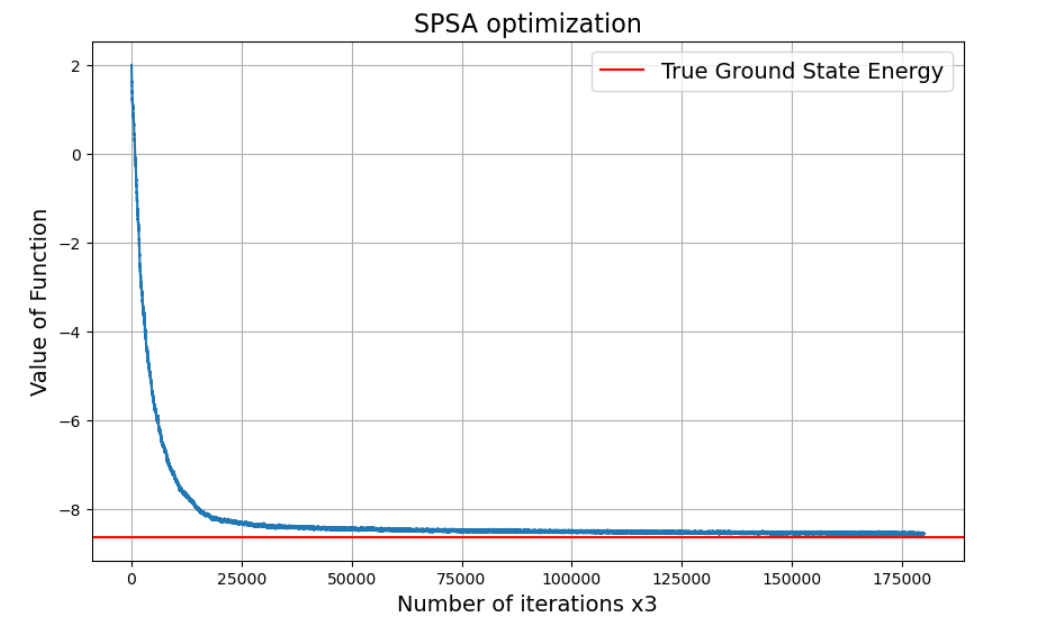}
    \caption{This figure illustrates SPSA iteratively computing and converging to the true ground state energy for the 2x4 U=0 Case. This corresponded to a 5-layer Big Ansatz trial. }
    \label{fig:enter-label}
\end{figure}

\newpage

\textbf{Simulated Ground State Energy Ansatz Layer Dependence Results (1x4, 2x2)}\\

Shown below in Table II as well as its corresponding figures Fig. 20 and Fig. 21 are the simulated ground state energy results that demonstrate the limitations of an optimization using an insufficient number of total parameters and illustrate the process of improvement from configuring the optimization with additional ansatz layers. Each ansatz layer is maximally parameterized with 16 parameters for the 1x4 lattice and 20 parameters for the 2x2 lattice. \\

\begin{table}[H]
\centering
\ra{ 1.3 }
\begin{tabular}{|c|c|c|c|c|c|c|c|} 
\toprule
Lattice & $U$ & True $E_{g}$ & $E_{Simulated}$ & \% Error & Ansatz Layers & Iterations\\
\midrule
1x4 & 0 & -2$\sqrt{5}$ & -4.3191 & 3.4\% & 1 & 2000 \\
1x4 & 0 & -2$\sqrt{5}$ & -4.4707 & 0.03\% & 2 & 2000 \\
\midrule
1x4 & 1 & -3.5753 & -3.53532 & 1.12\% & 2 & 2000 \\
1x4 & 1 & -3.5753 & -3.5547 & 0.58\% & 3 & 2000 \\
\midrule
2x2 & 0 & -4.00000 & -3.99697 & 0.08\% & 1 & 1000 \\
\midrule
2x2 & 1 & -3.3408 & -3.08502 & 7.66\% & 1 & 6000 \\
2x2 & 1 & -3.3408 & -3.26017 & 2.41\% & 2 & 6000 \\
2x2 & 1 & -3.3408 & -3.2884 & 1.57\% & 3 & 6000 \\
2x2 & 1 & -3.3408 & -3.2918 & 1.47\% & 4 & 6000 \\
\bottomrule
\end{tabular}
\caption{Ground state energy $E_{Simulated}$ results of studied lattices for different values of $U$. Hopping amplitude $t$ = 1 for all configurations. Results shown are best out of 20 SPSA optimization runs for each configuration. This table shows results of each case with varying number of ansatz layers except for the 2x2 U=0 case where no additional layers were required.}
\end{table}

\textbf{1x4 Lattice $E_{Simulated}$ Ansatz Layer Dependence For U=0 and U=1 Cases}
\begin{figure}[H]
\centering
\includegraphics[width=.7\linewidth]{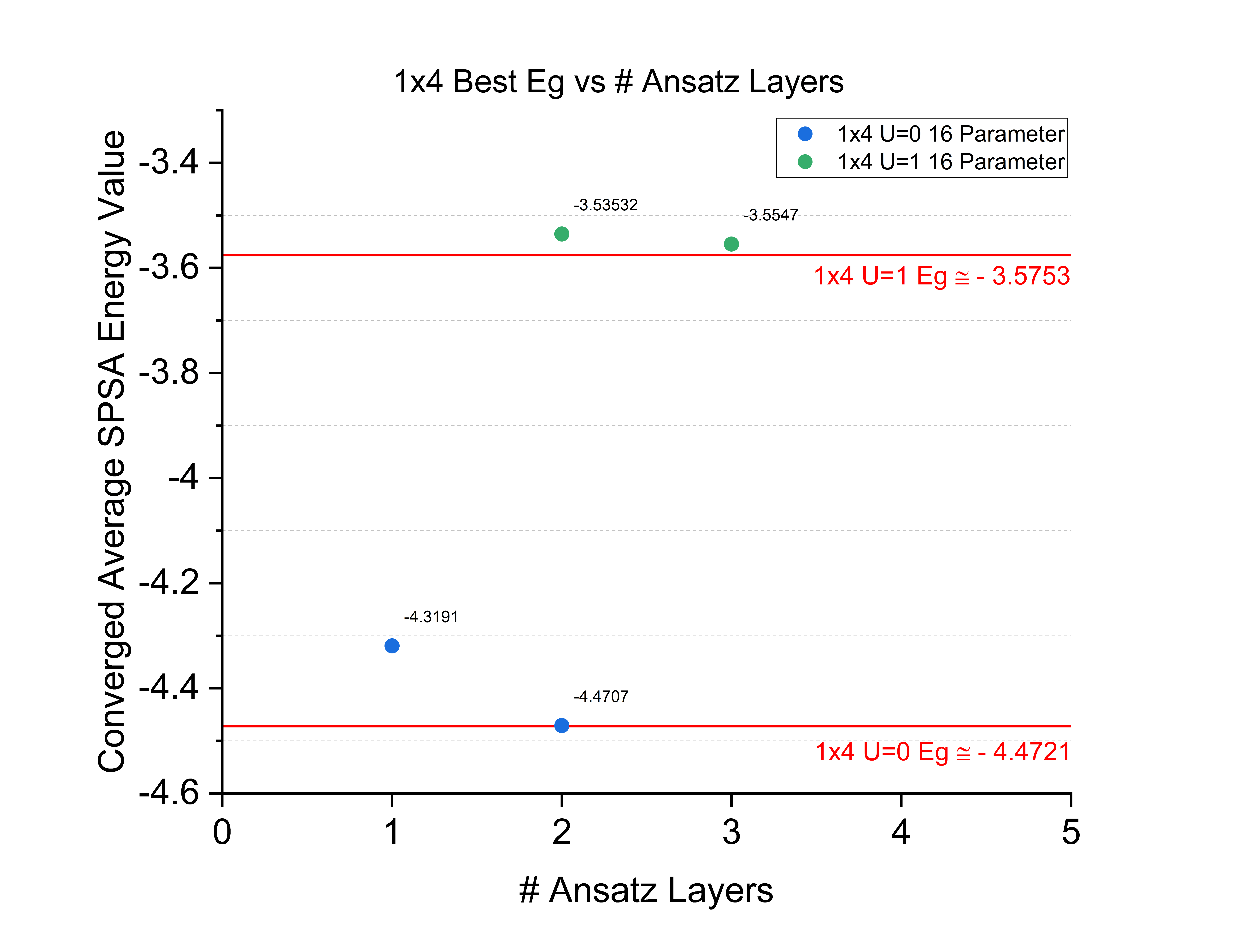}
\caption{This figure illustrates 1x4 lattice $E_{Simulated}$ results as a function of the number of ansatz layers using the maximally parameterized ansatz for the 1x4 lattice of 16 parameters per layer.}
\end{figure}
\newpage
\textbf{2x2 Lattice $E_{Simulated}$ Ansatz Layer Dependence For U=0 and U=1 Cases}
\begin{figure}[H]
\centering
\includegraphics[width=.7\linewidth]{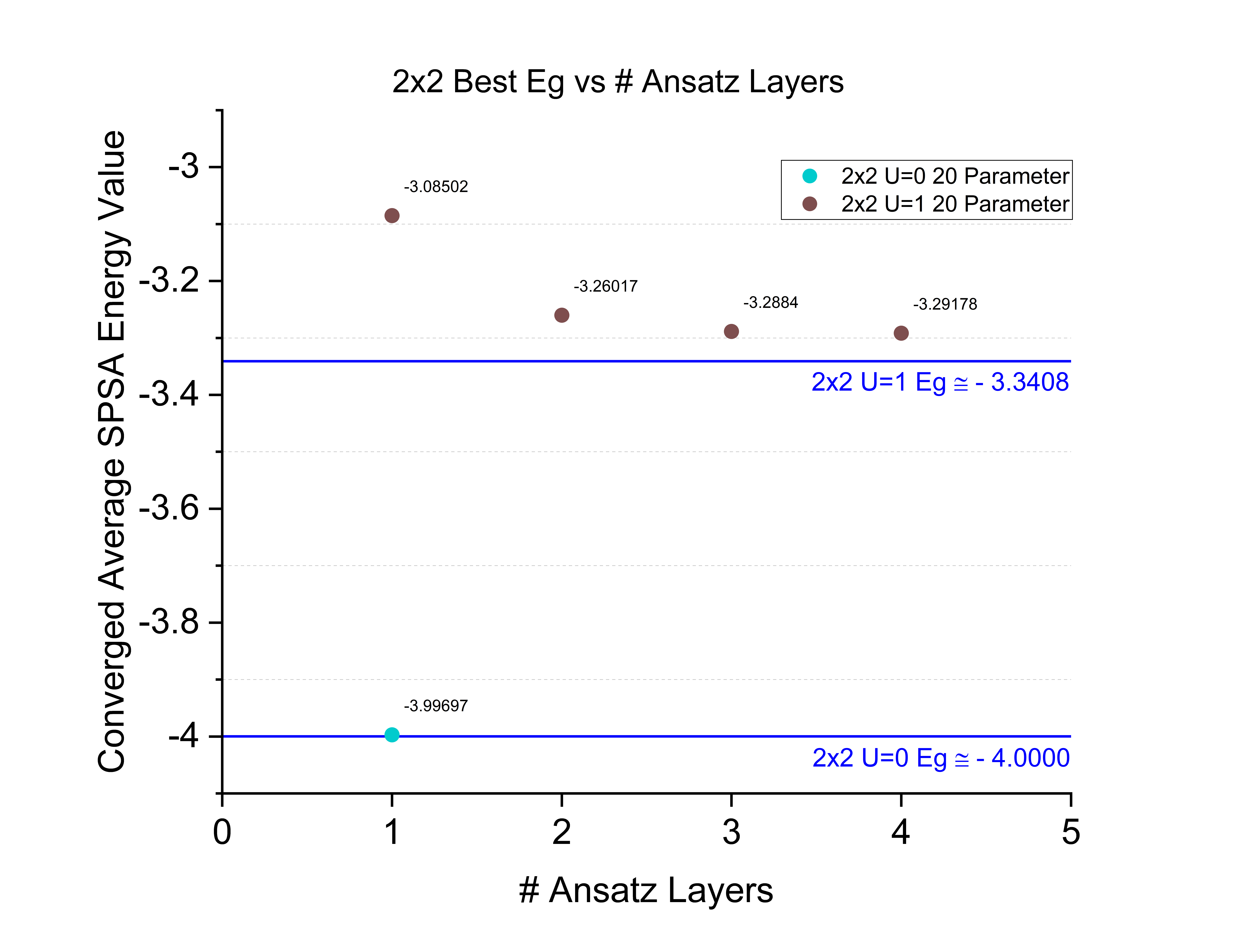}
\caption{This figure illustrates 2x2 lattice $E_{Simulated}$ results as a function of the number of ansatz layers using the maximally parameterized ansatz for the 2x2 lattice of 20 parameters per layer.}
\end{figure}

\newpage

\textbf{Simulated Ground State Energy Ansatz Layer Dependence Results (2x4 U=0)}
\begin{table}[H]
\centering
\ra{ 1.3 }
\begin{tabular}{|c|c|c|c|c|c|c|c|} 
\toprule
Lattice & $U$ & $True E_{G}$ &$E_{Simulated}$ & \% Error & Ansatz Layers & Ansatz Size & Iterations\\
\midrule
2x4 & 0 & -10.47 & -9.101 & 13.1\% & 1 & Large & 11000\\
2x4 & 0 & -10.47 & -8.091 & 22.7\% & 1 & Small & 11000\\
2x4 & 0 & -10.47 & -10.153 & 3.02\% & 2 & Large & 20000\\
2x4 & 0 & -10.47 & -9.77 & 6.69\% & 2 & Small & 20000\\
2x4 & 0 & -10.47 & -10.451 & 0.18\% & 3 & Large & 50000\\
2x4 & 0 & -10.47 & -10.016 & 4.34\% & 3 & Small & 50000\\
2x4 & 0 & -10.47 & -10.442 & 0.26\% & 4 & Large & 50000\\
2x4 & 0 & -10.47 & -10.172 & 2.84\% & 4 & Small & 50000\\
2x4 & 0 & -10.47 & -10.444 & 0.25\% & 5 & Large & 50000\\
2x4 & 0 & -10.47 & -10.089 & 3.64\% & 5 & Small & 50000\\
\bottomrule
\end{tabular}
\caption{Ground state energy $E_{Simulated}$ results of studied lattices for $U=0$. Hopping amplitude $t$ = 1 for all configurations. The amount of trials ranged from 10-20 depending on the layer and U case. This table shows results of each case with varying number of ansatz layers.}
\end{table}

Shown above in Table III are the simulated ground state
energy results demonstrating the effects and illustrate the process of improvement from configuring the optimization with additional ansatz layers. The 2x4 U=0 Small Ansatz requires 49 parameters, while the 2x4 U=0 big ansatz requires 117 parameters. We can see that the big ansatz gives a percent error that is smaller than that for the small ansatz by a factor of 2 or greater. The differences in the energy eigenvalues are negligible for the small ansatz between the 4 and 5 layers, again confirming our analysis that a 3-layer approach is most likely the most optimal when extending the algorithm to larger lattice sizes. The 3-layer small ansatz percent error of 0.18\% is the lowest, proving the accuracy of our algorithm and that this is a feasible way to use quantum computers in computing the ground state energy of the Fermi-Hubbard model. Many more iterations were required due to the greater circuit sizes in order for the optimizer to properly converge. A more convenient visualization for the U=0 case is shown below in Figure 22.

\begin{figure}[ht!]
    \centering
    \includegraphics[width=0.7\linewidth]{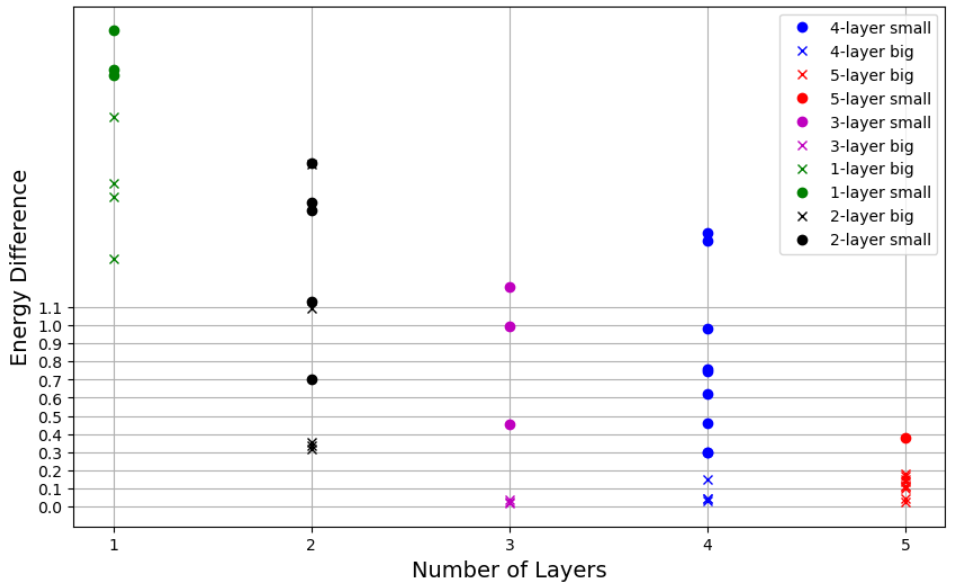}
    \caption{This figure illustrates SPSA an overview of the trials for the 2x4 U=0 Case. The number of trials ran were reduced from the smaller lattice size as each circuit required more time to run and create. }
    \label{fig:enter-label}
\end{figure}

As seen in the plot legend, the \textit{X} points correspond to big ansatz trials while the dots correspond to small ansatz trials. The y-axis corresponds to the energy difference between the true value (determined classically to be -10.47) and the value that the optimizer converged at. Due to the stochastic nature of the optimizer, the ground state energy results vary from run to run, so multiple trials are needed to find the correct ground state energy value. More layers can be used to improve the accuracy of the algorithm as it would allow for more degrees of freedom and parameters with which to control the gates and minimize the energy. In addition, both \textit{small} and \textit{big} ansatz's were plotted. A small ansatz is no different than those used in the previous cases, while a big ansatz is more rigorous in enabling communication between all qubits. This allows for more communication and degrees of freedom which improves accuracy. 

A general trend from the scatter plot is that increasing the number of layers decreases the energy difference, which matches our predictions in reducing error. This trend, however, stops after the 3rd layer as the difference becomes negligible. This suggests that more layers may not be needed if there are 3-5 layers, although further testing must be done with larger lattice sites. It is important to find an optimal configuration of which ansatz and how many layers to use in order to maximize efficiency in running simulations. The graph indicates that 3 layers combined with a large ansatz are optimal in running larger simulations, as it has an energy difference that is lower than or negligible compared to the additional layer trials. We can also see that the smaller ansatz configurations almost always result in an energy difference, and therefore percent error, that is higher than their big ansatz counterparts. This confirms our prediction that with more degrees of freedom, the optimizer can converge upon a value close to the true value. The table below shows the parameters relating to the best run for each ansatz and layer.\\

\textbf{2x4 Lattice U=1 Discussion and Limitations}\\
\begin{table}[H]
\centering
\ra{ 1.3 }
\begin{tabular}{|c|c|c|c|c|c|c} 
\toprule
Lattice & $U$ &$E_{Simulated}$ & Energy Difference & Ansatz Layers & Ansatz Size\\
\midrule
2x4 & 1 & -7.834 & 0.796 & 4 & Small\\
2x4 & 1 & -7.423 & 1.207 & 3 & Small\\
2x4 & 1 & -7.272 & 1.365 & 5 & Large\\
\bottomrule
\end{tabular}
\caption{$E_G$ = -8.63. Best Ground state energy $E_{Simulated}$ results for varying lattice sizes with U=1. Hopping amplitude $t$ = 1 for all configurations. The amount of trials ranged from 10-20 depending on the layer and U case.}
\end{table}

For the 2x4 U=1 case, $E_G$ = -8.63. The energy difference was calculated to be $|E_G| - |E_{Simulated}|$. Not all trials converged properly as there is a chance for the converging algorithm to find a "local" minimum (usually around 0). Once this happens, the converger sticks to this point which  requires the trial to be restarted. These error in these results are evidently much higher than their U=0 counterparts which can partly be attributed to the lack of computing resources. As seen below in 3-layer small ansatz trial, the code stopped running before the SPSA optimizer fully converged. This is due to the 72 hour time limit on jobs for our computing cluster. With more resources, a more accurate value can be obtained.

Higher values of $U$ (U=10) were also tested which gave unreasonable percent errors. This suggests that improvements need to be made to the Ansatz and initialization parts of the circuit. Utilizing Qiskit's "statevector initialize" feature can allow us to improve the initialization circuit by adding more of the possible basis states. Adding further CRY gates in the Ansatz may as well lead to improvements. This also may suggest that the optimizer is not a suitable way to calculate the ground state energies with high Coloumb repulsion, but more testing needs to be done to ensure this.\\

Shown in Table IV as well as Fig. 23 is the simulated ground state energy results that demonstrate the limitations of an optimization using an insufficient number of parameters per ansatz layer and illustrate the process of improvement from configuring the optimization with additional parameters for a given number of layers. \\
\newpage
\begin{figure}[ht!]
    \centering
    \includegraphics[width=.65\linewidth]{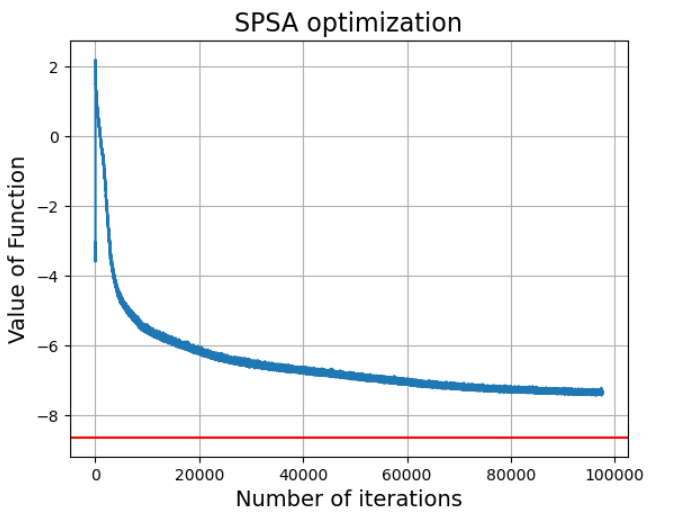}
    \caption{2x4 U=1 3-layer with small Ansatz. A more accurate value can be obtained since the graph shows that the optimizer is still in the process of converging.}
    \label{fig:enter-label}
\end{figure}

\textbf{Simulated $E_G$ Results Including Non-Maximally Parameterized Ansatz Configurations}\\
\begin{table}[H]
\centering
\ra{ 1.3 }
\begin{tabular}{|c|c|c|c|c|c|c|c|} 
\toprule
Lattice & $U$ & True $E_{g}$ & $E_{Simulated}$ & \% Error & Ansatz Layers & Parameters/Layer & Iterations \\
\midrule
1x4 & 0 & -2$\sqrt{5}$ & -4.1700 & 6.6\% & 1 & 3 & 2000 \\
1x4 & 0 & -2$\sqrt{5}$ & -4.2470 & 5.0\% & 1 & 5 & 2000 \\
1x4 & 0 & -2$\sqrt{5}$ & -4.3191 & 3.4\% & 1 & 16 & 2000 \\
1x4 & 0 & -2$\sqrt{5}$ & -4.33397 & 3.1\% & 2 & 3 & 2000 \\
1x4 & 0 & -2$\sqrt{5}$ & -4.4707 & 0.03\% & 2 & 16 & 2000 \\
\midrule
2x2 & 0 & -4.00000 & -3.99932 & 0.017\% & 1 & 7 & 1000 \\
2x2 & 0 & -4.00000 & -3.99697 & 0.08\% & 1 & 20 & 1000 \\
\midrule
2x2 & 1 & -3.3408 & -3.0300 & 9.3\% & 1 & 7 & 2000 \\
2x2 & 1 & -3.3408 & -3.0850 & 7.7\% & 1 & 20 & 2000 \\
2x2 & 1 & -3.3408 & -3.1018 & 7.2\% & 2 & 7 & 2000 \\
2x2 & 1 & -3.3408 & -3.2602 & 2.4\% & 2 & 20 & 2000 \\

\bottomrule
\end{tabular}
\caption{Ground state energy $E_{Simulated}$ results of studied lattices for different values of $U$. Hopping amplitude $t$ = 1 for all configurations. This table shows differences in performance for ansatz with different numbers or parameters per layer also shown below in Fig. 24.}
\end{table}
\newpage
\textbf{1x4 and 2x2 Lattice Ansatz Performance Using Different Numbers of Parameters/Layer}
\begin{figure}[H]
\centering
\includegraphics[width=.7\linewidth]{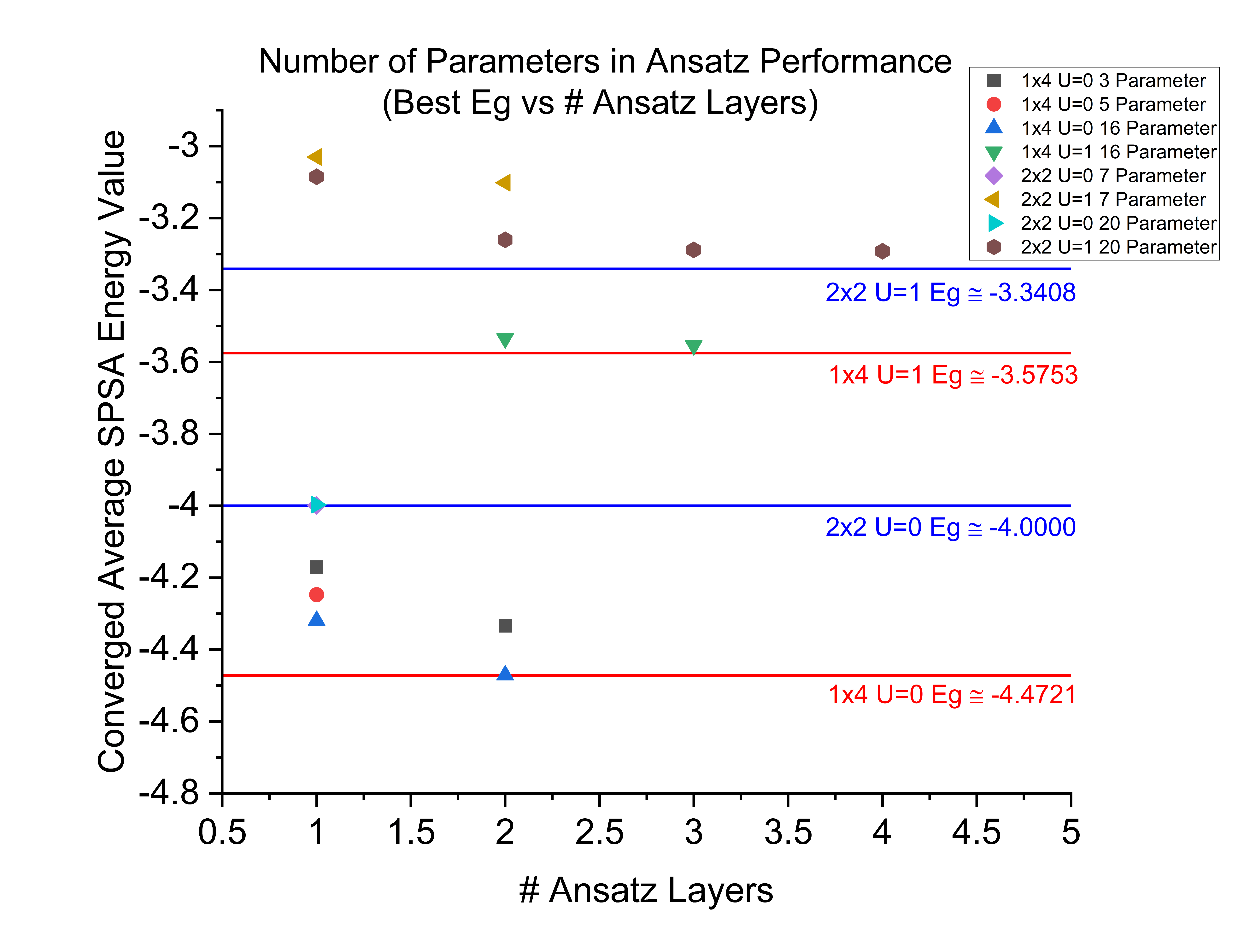}
\caption{This figure illustrates the best obtained 1x4 and 2x2 lattice $E_{Simulated}$ results using different numbers of parameters per ansatz layer as a function of the number of ansatz layers. These results are also shown in Table III above. The maximally parameterized ansatz performs better as shown by their lower computed ground state energies. The exception to this is when fewer parameters per layer is sufficient to accurately compute the ground state energy. In this situation as shown by the 2x2 $U=0$ case, there is no performance benefit to maximizing the number parameters within an ansatz layer (7 parameter result in purple is behind 20 parameter result in cyan).}
\end{figure}

\textbf{3x4 Implementation and Limitations}

\begin{table}[ht!]
\centering
\ra{ 1.3 }
\begin{tabular}{|c|c|c|c|c|c|c|c|} 
\toprule
Ansatz Layers & Ansatz Size & Number of Parameters\\
\midrule
2 & Large & 597\\
\midrule
3 & Large & 893\\
3 & Small & 295\\
\midrule
4 & Large & 1111\\
4 & Small & 391\\
\bottomrule
\end{tabular}
\caption{Parameter and circuit sizes for various circuits tested for the 3x4 U=0 Case. Construction of higher order circuits took longer amounts of time, hence the relatively smaller amount tested compared to the smaller lattice cases.}
\end{table}

The exponential increase in circuit size from higher lattice structures can be seen in the transition from the 2x4 to 3x4 lattice site. The table above demonstrates this. Evidently, the number of parameters increases significantly which makes constructing larger lattice sites more difficult. The 4-layer large ansatz model had 1111 parameters and the 5-layer would likely require around or greater than 1400 parameters. This circuit was initialized according to the previously mentioned  procedures and followed the same format and restrictions. 

Starting from the observation that the 3-layer provided an optimal result for the 2x4 U=0 case, we first constructed the 3x4 U=0 case with 3-layers as well. We found that our results were limited by the amount of computing resources available to our lab. These circuits required approximately 5-7 days to run; this is expected as each of the 7 circuits is run 40000 times and calculations are done on the physical states, after which the optimizer attempts to converge and this process is repeated 100,000 times. The maximum amount of time we could run our files for was 72 hours, which fell significantly below the necessary time frame. We experimented with the 4 layer circuit of both ansatz sizes to see if a difference could be made but, as expected, the circuit was not able to converge in time. The same applied to the 2-layer circuit. It is clear that even if the 1-layer provided a suitable convergence, the ground state energy would not be correct as multiple layers are needed for accuracy due to the increased degrees of freedom. 

\begin{figure}[ht!]
    \centering
    \includegraphics[width=0.7\linewidth]{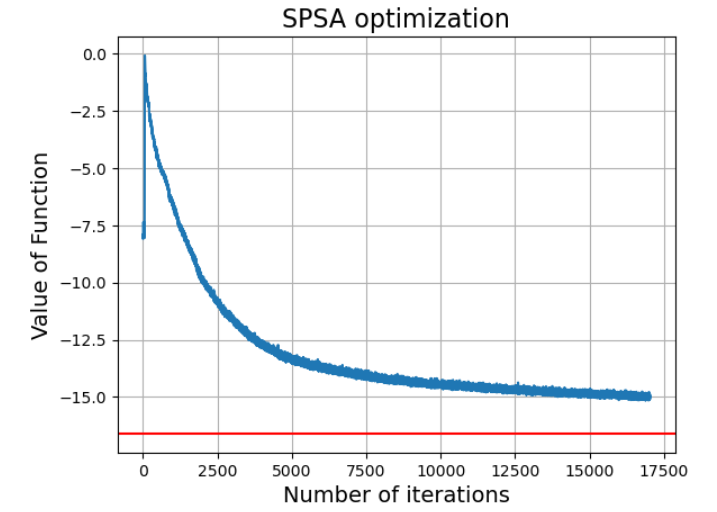}
    \caption{This figure illustrates SPSA iteratively computing and converging to the true ground state energy for the 3x4 U=0 case. This was done using 3 Ansatz Layers with a parameter size of 893.}
    \label{fig:enter-label}
\end{figure}

Figure 24 represents one of the best 3x4 trial runs with the current hardware limitations. The true $E_G$ value of the 3x4 U=0 lattice is -16.6 and the optimizer left off at an average value of approximately -15.0028, which gives an energy difference of 1.5971. We cannot determine with certainty where this graph will converge due to the stochastic nature of the optimizer, but seeing the convergence behavior appearing close to the true value (computed classicaly), it is reasonable to assume that with more computing time, this will converge close to the true value. Therefore, this algorithm is a viable option for further research. 

\section{conclusion}

These results that are in close agreement with the true values suggest a successful method has been established for simulating small Fermi-Hubbard spin lattices where accuracy was determined within 0.6\% error in all cases except the 2x2 lattice U=1 case within 1.5\% error. This case will be investigated for further optimization implementing techniques including adding more ansatz layers, allowing the optimization to run for significantly more iterations to allow the optimizer to explore a potentially large plateau, or evolving the state of a higher value of $U$ such as $U=10$ or $U=100$ towards $U=1$ by recursively using optimized parameters of the previous case as initial parameters for the next. For the larger 2x4 lattice size, the true energy value was determined within 0.18\% error for the 3-layer large ansatz case. The 4-layer and 5-layer circuit determined the true value with the largest error being 3.64\%. For the U=1 case, the closest energy difference was 0.796 which corresponded to a 4-layer small ansatz. This model, along with the 3x4 lattice, can be improved with more computing time and the configuration of the initialization state as mentioned before. Due to the accurate results, many more configurations can be studied in the future using this method, potentially within this research group. See the future work section below for more details.\\

\newpage
\section{Future Work}

\textbf{Improve 2x4, 3x4 cases}\\

Improvement of the 2x4 U=1, 3x4 U=0 case will be investigated with multiple methods such as state evolution, additional ansatz layers, and more iterations as mentioned in the conclusion. \\

\textbf{Larger lattices: Ground state energy computations} \\

Construction and simulation of larger lattices including the 4x4 will be further studied. Additionally, increasing the Coloumb repulsion (U-values) will be investigated to see the effectiveness of the model. \\

\textbf{Generalization of initialization circuits for larger lattices}\\

Generalizing scalable initialization circuits for desired occupancy levels for lattices of odd and even dimensions (separate tasks) that produce a superposition of as many physical basis states as possible is important for alleviating the optimization workload. I.e., the optimizer has to work harder with fewer initial physical basis states and potentially must work harder with non-uniform physical basis state amplitudes. \\


\textbf{Run on a real quantum computer}\\

The research group encourages testing computing ground state energies of lattices in this study as well as larger lattices on a real quantum machine instead of simulated quantum circuits. \\

\textbf{Noise mitigation}\\

Effects of noise and noise mitigation have studied within the research group by Nidish Narsipur. Some of Nidish's methods include error multiplication via multiplying the number of rotation gates in the quantum circuit and reducing rotation angles proportionally such that the same state is prepared with a multiplication of noise associated with each additional gate. These multiples of error amplification are plotted and polynomial regression is used to determine accurate results despite the presence of noise. So far this method has been successfully run on a real quantum computer for the 1x2 lattice.\\

\newpage

\bibliography{citations}

\begin{thebibliography}{1}

\bibitem{paper6}
Charles~Derby et~al.
\newblock Efficient and practical hamiltonian simulation from time-dependent product formulas.
\newblock {\em ArXiV}, (08729), 2024.

\bibitem{quantumbook}
Ramamurti Shankar.
\newblock {\em Principles of Quantum Mechanics}, volume~1.
\newblock Plenum Publishers, 1st edition, 1994.

\bibitem{paper2}
Chris~Cade et~al.
\newblock Strategies for solving the fermi-hubbard model on near-term quantum computers.
\newblock {\em Physical Review B}, 102(235122), 2020.

\bibitem{qiskitbook}
Russell~Huffman et~al.
\newblock {\em IBM Quantum Qiskit Textbook}, volume~1.
\newblock IBM Research, Qiskit Community, 1st edition, 2017.

\bibitem{rieffel}
Wolfgang~Polak Eleanor~Rieffel.
\newblock {\em Quantum Computing: A Gentle Introduction}, volume~1.
\newblock The MIT Press, 1st edition, 2014.

\bibitem{paper4}
Xiqiao~Wang et~al.
\newblock Experimental realization of an extended fermi-hubbard model using a 2d lattice of dopant-based quantum dots.
\newblock {\em Nature Communications}, (6824), 2021.

\bibitem{paper1}
Daniel P.~Arovas et~al.
\newblock The hubbard model.
\newblock {\em Annual Reviews}, 13:239-274, 2021.

\bibitem{paper5}
Chengfang~Cao et~al.
\newblock Unveiling quantum phase transitions from traps in variational quantum algorithms.
\newblock {\em ArXiV}, (08441), 2024.

\bibitem{paper3}
Stasja~Stanisic et~al.
\newblock Observing ground-state properties of the fermi-hubbard model using a scalable algorithm on a quantum computer.
\newblock {\em Nature}, 13(5743), 2022.

\end{thebibliography}
\bibliographystyle{unsrt}

\end{document}